\documentclass[epj]{svjour}

\usepackage{graphicx}
\usepackage{fancyhdr}
\def\makeheadbox{{%  remove EPJC header
 }}
\def\mytitle{My title} 
\def\myauthors{My name}  
\def\mytype{My type of session}
\def\mysession{My session}
\def\mytitle{Supersymmetry Breaking
by Constant Boundary Superpotentials in Warped Space}%Short title of talk} %Put your title here!
\def\myauthors{Nobuhiro Uekusa} %Name of Author}    %Put your name here!
\def\mytype{Parallel Session Talk} %Contributed Talk}    
\def\mysession{Alternatives}
\newcommand{\ccr}{c_{\textrm{\scriptsize cr}}}
\begin{document}
\title{Supersymmetry Breaking and Radius Stabilization \\
by Constant Boundary Superpotentials 
in Warped Space}
\author{Nobuhito Maru\inst{1}
\and Norisuke Sakai\inst{2}
\and Nobuhiro Uekusa\inst{3}
\thanks{\emph{Email:} nobuhiro.uekusa@helsinki.fi} 
}                     % Do not remove
\institute{
Department of Physics, Kobe University, 
Kobe 657-8501, Japan
\and
Department of Physics, Tokyo Institute of Technology, 
Tokyo 152-8551, Japan
\and
High Energy Physics Division, 
Department of Physical Sciences, University of Helsinki  
and Helsinki Institute of Physics, 
P.O. Box 64, FIN-00014 Helsinki, Finland}
\date{}
\abstract{
Supersymmetry breaking and radius stabilization 
by constant superpotentials localized 
at boundaries is studied 
in a supersymmetric warped space model where 
a hypermultiplet, a compensator and a radion multiplet
are taken into account.
Soft mass induced by the 
anomaly mediation can be of the order of 100GeV and can 
be dominant compared to that mediated by bulk fields. 
A lighter physical mode composed of 
the radion and the moduli can have mass of the order of 
a TeV and the gravitino mass can be of the order of 
10$^7$ GeV. 
The radius is stabilized by the presence of 
the constant boundary superpotentials.
We also find that the mass splitting has an interesting dependence on 
the bulk mass parameter $c$. 
\PACS{
{11.30.Pb}{Supersymmetry} \and
{11.25.-w}{Strings and branes} \and
{12.60.Jv}{Supersymmetric models}
     } % end of PACS codes
} %end of abstract
\maketitle
\section{Introduction}
\label{intro}
Supersymmetry is a well-motivated extension to the standard 
model, which plays a crucial role in solving the gauge 
hierarchy problem \cite{Dimopoulos:1981zb}. 
Extra dimensions are also an 
alternative solution to the gauge hierarchy problem 
%\cite{Arkani-Hamed:1998rs} \cite{Randall:1999ee}
\cite{Arkani-Hamed:1998rs}. 
Considering both ingredients is natural in the context 
of the string theory and is often taken as the starting 
point in the phenomenological model of the brane world 
scenarios. 
In such a setup, we have to compactify extra dimensions 
and break supersymmetry to obtain realistic four-dimensional 
physics.
One of the simple ways to realize it is the Scherk-Schwarz 
mechanism of supersymmetry breaking \cite{SS}. 
It is known that the Scherk-Schwarz supersymmetry breaking 
is equivalent 
to the supersymmetry breaking by a constant 
superpotential 
in flat space %\cite{MP,BFZ,GR}.
\cite{MP,BFZ}
It is natural to ask whether this equivalence still holds 
in warped space. This issue has been discussed in the literature 
%\cite{HNOO, BB, AS}.   \cite{ABN, GP, FLP, BKV}
\cite{Gherghetta:2000qt,HNOO}.

We present a brief summary of our study
of super- symmetry-breaking effects and radius stabilization 
in a warped model with supersymmetry broken by
constant boundary superpotentials \cite{Maru:2006id,MSU}.
Taking the hypermultiplet and including the compensating 
multiplet and the radion multiplet, 
we show that the radius is stabilized by the presence of 
the constant boundary superpotentials.
It is also found that the mass spectrum depends on the bulk mass 
parameter in addition to the strength of the 
constant boundary superpotential. 
A lighter physical mode composed of 
the radion and the moduli can have masses of the order of 
a TeV and that the gravitino mass can be of the order of 
10$^7$ GeV.
It is also shown that induced mass mediated by 
anomaly can be of the order of 100GeV and can be dominant 
compared to that mediated by bulk fields. 
%We find that there is no flavor-changing neutral-current problem 
%in a wide range of 
%parameters.

\section{Model} 
\label{sec:1} 
We consider a five-dimensional supersymmetric model of 
a single hypermultiplet on the 
Randall-Sundrum 
background, 
whose metric is 
\begin{eqnarray}
ds^2 = e^{-2R\sigma}\eta_{\mu\nu}dx^\mu dx^\nu +R^2 dy^2, 
\quad 
\sigma(y)\equiv k|y|, 
\end{eqnarray}
where $\eta_{\mu\nu}={\rm diag.}(-1,+1,+1,+1)$, 
$R$ is the radius of $S^1$ of the orbifold $S^1/Z_2$, 
$k$ is the $AdS_5$ curvature scale, and the angle of $S^1$ 
is denoted by $y(0 \le y \le \pi)$. 
In terms of superfields for four manifest supersymmetry, 
our Lagrangian reads \cite{MP}
\begin{eqnarray}
{\cal L}_5 &=& \int d^4 \theta 
\frac{1}{2} \varphi^\dag \varphi (T+T^\dag) 
e^{-(T+T^\dag)\sigma} 
\\
\nonumber
&&
\times
(\Phi^\dag \Phi + \Phi^c \Phi^{c\dag} - 6M_5^3) 
\nonumber \\
&& + \int d^2 \theta 
\left[
\varphi^3 e^{-3T \sigma} \bigg\{
\Phi^c \left[
\partial_y - \left( \frac{3}{2} - c \right)T \sigma' 
\right] \Phi
\right. 
\nonumber
\\
&&\left.
+ W_b
\bigg\} + {\rm h.c.}
\right] ,
\label{lagrangian}
\end{eqnarray}
where the compensator chiral supermultiplet $\varphi$ 
(of supergravity), and the radion chiral supermultiplet 
$T$ are denoted as
$\varphi = 1 + \theta^2 F_{\varphi}$ and 
$T=R + \theta^2 F_T$,
respectively, and the chiral supermultiplets representing 
the hypermultiplet is denoted as 
$\Phi, \Phi^c$.
The $Z_2$ parity is assigned to be even (odd) for 
$\Phi (\Phi^c)$. 
The derivative with respect to $y$ is denoted by $'$, 
such as $\sigma'\equiv d\sigma/dy$. 
The five-dimensional Planck mass is denoted as $M_5$. 
Here we consider a model with a constant (field independent) 
superpotential localized at the fixed point $y=0$ 
\begin{eqnarray}
W_b \equiv 2M_5^3 w_0 \delta(y) ,
\label{eq:boundary_pot}
\end{eqnarray}
where $w_{0}$ is a dimensionless constant.

\section{Radius stabilization}
\label{sec:2} 
\subsection{Background solutions}

The background solutions for the scalar components 
at the leading order of $w_0$ are given by
\begin{eqnarray}
%&& 
\phi(y)&=&N_2\exp\left[\left({3\over 2}-c\right)R\sigma\right], 
 \label{phin2}
\\ 
\phi^c(y)&=&\hat{\epsilon}(y)
   \left({\phi^\dagger\phi\over 6M_5^3}-1\right)^{-1}
 \left({\phi^\dagger\phi\over 6M_5^3}\right)^{{5/2-c\over 3-2c}}
\nonumber
\\
&& \times
   \left[c_1 +c_2 
 \left({\phi^\dagger\phi\over 6M_5^3}\right)^{-{1-2c\over 3-2c}}
\left({\phi^\dagger\phi\over 6M_5^3}+{2\over 1-2c}\right)\right]
\nonumber
\\
\label{phichic}
\end{eqnarray}
where $c\neq 1/2, 3/2$, and 
\begin{eqnarray}
\hat{\epsilon}(y) \equiv 
\left\{
\begin{array}{cc}
+1, & 0<y<\pi \\
-1, & -\pi<y<0
\end{array}
\right.
. 
\label{eq:sign_function}
\end{eqnarray}
The solution contains three complex integration constants: 
$c_1, c_2$ are the coefficients of two independent 
solutions for $\phi^c$, and the overall complex constant 
$N_2$ for the flat direction $\phi$. 
Two out of these three complex integration constants 
are determined by the boundary conditions. 
The single remaining constant (which we choose as $N_2$) 
is determined through the minimization of the potential 
(stabilization). 

\subsection{Potential}
\begin{figure}[htbp]
\begin{center}
\includegraphics[width=0.3\textwidth,angle=0]{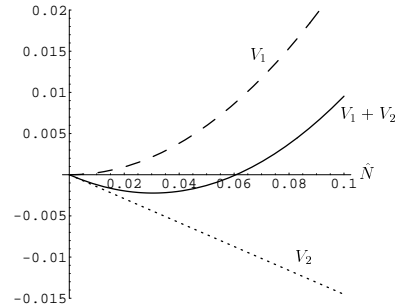}
\caption{Potential for $c=c_{\textrm{\scriptsize cr}}-\Delta c$}
\label{fig:1}
\end{center}
\end{figure}

With the backgrounds (\ref{phin2}) and (\ref{phichic}),
the potential is obtained as 
\begin{eqnarray}
 V
&=& {3M_5^3 k w_0^2\over 2}
\nonumber
\\
 &&\times
\bigg\{
  \frac{-2(1-2c)}{(1-2c)(e^{2Rk\pi}-1)\hat{N} 
+2(e^{(2c-1)Rk\pi}-1)}
\nonumber
\\
&&\times
\hat{N}^{4-2c-\frac{1}{3-2c}} \nonumber \\
&&+\frac{\hat{N}}{1-\hat{N}}
\left( -4c^2+12c-6 +\frac{3-2c}{3(1-\hat{N})}
\right)
  \bigg\}. 
\label{potentialwp0}
\end{eqnarray}
where $\hat{N}\equiv |N_2|^2/(6M_5^3)$.
We need to require the stationary condition for both modes $R$ and
$N_2$,
$\partial V/\partial R=0$ and 
$\partial V/\partial \hat{N}=0$.
From the stationary condition,
we find that there is a unique nontrivial minimum with a finite 
value of the radius $R$ and the normalization $N_2$ for the 
flat direction $\phi$ provided 
$c < c_{\rm cr}$ with 
\begin{eqnarray}
  c_{\textrm{\scriptsize cr}}\equiv
    {17-\sqrt{109}\over 12} .
\end{eqnarray}
At the critical value of the mass parameter $c_{\rm cr}$, 
the minimum occurs at infinite radius and vanishing 
normalization $N_2$, 
$\hat{N}(\ccr)=0$, $R(\ccr)=\infty$. 
To examine the stabilization for $c < c_{\rm cr}$ more closely, 
we parameterize $c=\ccr-\Delta c$ with a small $\Delta c$. 
After using the stationary condition solution
$\hat{N}=e^{-(3-2c)Rk\pi}$, 
we find that the potential (\ref{potentialwp0}) for 
$c=\ccr-\Delta c$ at the leading order of 
$\Delta c$ and $\hat{N}$ consists of two pieces 
\begin{eqnarray}
  V  &\approx& {3M_5^3 kw_0^2\over 2}(V_1 +V_2) ,
\\
  V_1&\equiv& {2(2\ccr-1)\over 3-2\ccr}
    \hat{N}^{4\ccr^2-12\ccr+10\over 3-2\ccr} , 
\\
  V_2&\equiv& -\hat{N}\left(-8\ccr+{34\over 3}\right)\Delta c 
. 
\end{eqnarray}
The potential $V$ and its pieces $V_1, V_2$ are depicted 
as a function of $\hat N$ 
in Fig.\ref{fig:1}. 
It is now obvious that a unique minimum occurs at 
finite values of $\hat N$ provided $\Delta c >0$ 
($c < c_{\rm cr}$) and that the minimum point approaches 
$\hat N \to 0$ as $\Delta c \to 0$ 
($c \to c_{\rm cr}$).  
Actually the Fig.\ref{fig:1} demonstrates only 
the stability along the direction of $\hat N$, 
after the other variable $R$ is eliminated by the 
stationary condition. %(\ref{dvr0}). 
We have checked that this minimum point gives 
a true minimum of the potential $V(R, \hat N)$ 
as a function of two variables, establishing the stability 
in both directions. 
The stationary point at 
the leading order of $\Delta c$ is obtained as 
\begin{eqnarray}
  R&\approx&
    {-1\over \left[2(1-\ccr)(3-2\ccr)+1\right]k\pi}
\nonumber
\\
   &&\times
   \ln \left[
     {(3-2\ccr)\left({17\over 3}-4\ccr\right)
     \over 2(2\ccr-1)\left(2-\ccr-{1-\ccr\over 3-2\ccr}\right)}
   \Delta c\right] 
\nonumber \\ 
 &\approx& {1\over 10k}\left(\ln {1\over \Delta c}-3.4\right) ,
\label{Rdeltac} 
\end{eqnarray}
which means that the radius is stabilized with the size 
of $R>1/k$ 
for $\Delta c<10^{-6}$. 

At the stationary point the potential becomes 
\begin{eqnarray}
  V&\approx& -10^{37}(kw_0)^2(\Delta c)^{1.2} .
  \label{eq:pst}
\end{eqnarray}
We can show that the cosmological constant
can be cancelled by an F term 
contribution and a D term contribution
 for supersymmetry breaking
and that the contributions of these sectors to 
the soft mass and gravitino mass are small.

\section{Mass spectrum}

\subsection{Soft mass by anomaly mediation}
In a supersymmetric Randall-Sundrum model,
anomaly -mediated scalar mass is given by
$\tilde{m}_{\textrm{\scriptsize AMSB}}
 \sim (g^2/ 16\pi^2)$ $\cdot$ $\langle F_\omega/\omega\rangle$
\cite{Luty:2002ff}.
Here the superfield $\omega$ is defined as 
a rescaled compensator multiplet $\omega=\varphi e^{-T\sigma}$
and we denoted its lowest component also as $\omega$, 
and $g$ is gauge coupling constant for visible sector fields. 
In our model, the anomaly-mediated scalar mass becomes
\begin{eqnarray}
  \tilde{m}_{\textrm{\scriptsize AMSB}}
 \sim {g^2\over 16\pi^2}  (F_\varphi-F_T\sigma)\bigg|_{y=\pi} .
\end{eqnarray}
Using the the hyperscalar background (\ref{phin2}), (\ref{phichic})
and the stationary condition,
 we obtain the anomaly-mediated scalar mass as
\begin{eqnarray}
 \tilde{m}_{\textrm{\scriptsize AMSB}}
  \sim  {\cal O}(10^{-4})\times g^2 k w_0
\end{eqnarray}
For $g^2 kw_0\sim 10^6$GeV,
we obtain
\begin{eqnarray}
   \tilde{m}_{\textrm{\scriptsize AMSB}}
 \sim  100 \textrm{GeV} ,
\end{eqnarray}
which is a typical soft mass.
We can show that soft masses by mediation of Kaluza-Klein modes
in our model are smaller than
that of anomaly mediation.
Therefore our model passes the flavor-changing neutral current
constraint also with respect to 
bulk field mediation while 
$\tilde{m}_{\textrm{\scriptsize AMSB}}\sim 100\textrm{GeV}$.

For gaugino mass, anomaly mediation is also 
dominant as long as additional interactions with gauge singlets
are not included.
The gaugino mass is of the same order as the scalar mass.

\subsection{Radion and moduli masses}
We calculate the masses for
the quantum fluctuations of the radion and moduli 
in our model.
%From Eq.(\ref{lagrangian}), the kinetic Lagrangian is 
%written as
%\begin{eqnarray}
% {\cal L}_{\textrm{\scriptsize kin}}
%&=& \sigma (\partial_\mu R)(\partial^\mu R)e^{-2R\sigma}
%  (\phi^\dagger\phi+\phi^c{}^\dagger\phi^c-6M_5^3 )
%\nonumber\\
%  &&-{1\over 2}(1-2R\sigma)
%e^{-2R\sigma}(\partial_\mu R)\partial^\mu
%   (|\phi|^2+|\phi^c|^2)
%\nonumber\\
% &&- Re^{-2R\sigma}\left[
%  (\partial_\mu
%  \phi^\dagger)(\partial^\mu \phi)
%  + (\partial_\mu
%  \phi^c{}^\dagger)(\partial^\mu \phi^c)\right] ,
%   \label{lag}
%\nonumber
%\\
%\end{eqnarray}
%where we performed a partial integral and dropped 
%four-dimensional total derivatives. 
%%
%Since the field $\phi^c$ is of higher order of $w_0$, 
%we will omit $\phi^c$ in Eq.(\ref{lag}). 
Without loss of generality, we can choose the phase of the 
background classical solution in Eq.(\ref{phin2}) as 
\begin{eqnarray}
N_2=N_2^\dagger. 
\end{eqnarray}
We now introduce quantum fluctuation fields around the 
background classical solution to define the radion 
$\tilde R$ and the moduli field $\tilde{N}_2$ : 
\begin{eqnarray}
R+\tilde{R}, \quad N_2+\tilde{N}_2, \quad 
\tilde{N}_2=\tilde{N}_{2R}+i\tilde{N}_{2I}. 
\end{eqnarray}
Substituting Eq.(\ref{phin2})
into the Lagrangian %${\cal L}_{\textrm{\scriptsize kin}}$
and diagonalizing the kinetic term and mass-squared matrix, 
we find that at the leading order of $e^{-Rk\pi}$
the lighter physical mode is almost exclusively 
made of the radion 
\begin{eqnarray}
 m^2_{\textrm{\scriptsize light
}}
% &\approx& k^2 w_0^2 
%   \frac{2(1-2c)^2}{3-2c}(Rk\pi)^2 e^{(-4c^2+12c-10)Rk\pi} 
%\nonumber
%\\
   &\approx& k^2 w_0^2 
  0.38 (3.4+ \ln \Delta c)^2 (\Delta c)^{1.7} . 
\end{eqnarray}
The heavier eigenmode is found to be exclusively made of 
the real part of moduli field 
\begin{eqnarray}
 m^2_{\textrm{\scriptsize heavy
}}
% &\approx& k^2 w_0^2 
% {(2c-1)\over 4}[-4c^2+12c-6+\frac{4}{3}(3-2c)]
%\nonumber
%\\
% &&\times
%{e^{(2c-3)Rk\pi} \over 1-e^{-(2c-1)Rk\pi}} 
%\nonumber
%\\
   &\approx& k^2 w_0^2 
  0.47 (\Delta c)^{0.70} . 
\end{eqnarray}
The imaginary part of the moduli field has the same mass as 
the real part of the moduli field in this approximation. 
We estimate the mass of the lighter physical mode (almost 
exclusively made of the radion), and that of the heavier 
mode (almost exclusively made of the 
complex moduli field) as 
\begin{eqnarray}
  m_{\textrm{\scriptsize light}} \sim  1 \textrm{TeV}, 
\qquad 
  m_{\textrm{\scriptsize heavy}} \sim 
 100 \textrm{TeV}
\end{eqnarray}
for $w_0\sim (10^{7}\textrm{GeV}/k)$ and $\Delta c\sim 10^{-6}$. 

\subsection{Gravitino mass}
The other superparticles affected by $w_0$ are gravitino and hyperscalar. 
The relevant gravitino Lagrangian in the bulk is given 
by \cite{Gherghetta:2000qt}
\begin{eqnarray}
{\cal L}_{{\rm bulk}} &=& 
 M_5\sqrt{-g}  \bigg[i\bar{\Psi}_M^i 
\gamma^{MNP} D_N \Psi_P^i
\nonumber
\\
&&  -\frac{3}{2} \sigma' 
\bar{\Psi}_M^i \gamma^{MN}(\sigma_3)^{ij} \Psi_N^j \bigg] ,
 \label{ginolag}
\end{eqnarray}
\begin{eqnarray}
\Psi_M^1 &=& (\psi_M^1{}_{\alpha}, 
            \bar{\psi}_M^2{}^{\dot{\alpha}})^T, ~~
\Psi_M^2 = (\psi_M^2{}_{\alpha}, 
            -\bar{\psi}_M^1{}^{\dot{\alpha}})^T, \\
D_M &=& \partial_M + \omega_M, ~~
\omega_M =(\omega_\mu, \omega_4) = (\sigma' \gamma_4 \gamma_\mu/2, 0), 
\nonumber
\\
\end{eqnarray}
%\begin{eqnarray}
%\gamma^{M_1 M_2\cdots M_N} &=&
%  \gamma^{[ M_1} \gamma^{M_2} \cdots \gamma^{M_N]} 
%\nonumber \\
% &\equiv& \frac{1}{N!}(\gamma^{M_1} \gamma^{M_2} \cdots \gamma^{M_N}
%\nonumber
%\\ 
%  && + \textrm{antisymmetric permutations}), 
%\end{eqnarray}
where the 5D curved indices are labelled by $M, N = 0, 1,2, 3, 4$.
The gamma matrix with curved indices is defined
through 5D vielbein as $\gamma^M=e_A^M\gamma^A$, 
where $A$ denote tangent space indices. 
In the second term in Eq.(\ref{ginolag}), 
$SU(2)_R$ indices are contracted by $(\sigma_3)$.

Boundary terms for gravitino are also contained 
in the term with the boundary superpotential $W_b$ 
in the superfield Lagrangian in Eq.(\ref{lagrangian}). 
By restoring the fermionic part, 
we find 
%\begin{eqnarray}
%{\cal L}_{\textrm{\scriptsize bound sup}} 
%= \int d^2\theta \varphi^3 e^{-3T\sigma} W_b
%= 3 \left[F_\varphi - \frac{1}{M_5^2}
%\psi^1_\mu\sigma^{[\mu}\bar \sigma^{\nu]}\psi^1_\nu 
%+ h.c.\right] W_b + \cdots
% ,
%  \label{eq:sup_bound_pot} 
%\end{eqnarray}
${\cal L}_{\textrm{\scriptsize bound sup}} 
= \int d^2\theta \varphi^3 e^{-3T\sigma} W_b
= 3 \left[F_\varphi - \frac{1}{M_5^2}
\psi^1_\mu\sigma^{[\mu}\bar \sigma^{\nu]}\psi^1_\nu 
+ \textrm{h.c.}\right] W_b + \cdots$,
using $\varphi=1+\theta^2F_\varphi$ \cite{Kugo:2002js}. 
Therefore we obtain a boundary mass term for gravitino 
associated to the boundary superpotential 
\begin{eqnarray}
{\cal L}_{{\rm boundary}} = 
 -\frac{3W_b}{M_5^2} 
  \left[\psi_{\mu}^1 \sigma^{[\mu}\bar{\sigma}^{\nu]} \psi_\nu^1 
      +\bar{\psi}_{\mu}^1 \bar{\sigma}^{[\mu}\sigma^{\nu]}
     \bar{\psi}_\nu^1 \right] .
  \label{ginolagb} 
\end{eqnarray}
Here we assumed the $Z_2$ parity of $\psi_\mu^{1(2)}$ to be even (odd). 

From the Lagrangian given above,
we calculate mass spectrum.
For the lightest mode
$m_n e^{Rk\pi} \ll k$, 
%\begin{equation}
%\frac{m_n}{k} \ll 1, ~~\frac{m_n}{k}e^{Rk\pi} \ll 1. 
%\end{equation}
we find 
\begin{eqnarray}
  m_{\textrm{\scriptsize lightest}} \approx 6w_0 k ,
  \label{gino}
\end{eqnarray}
which can be $10^7$GeV for $w_0 \sim (10^7\textrm{GeV}/k)$.
This shows that the four\,-\,dimensional gravitino (lightest mode) 
is much heavier than the 
the radion as well as scalars of the visible sector.
This is similar to the supersymmetry-breaking mediation model 
considered previously by Ref.\cite{Luty:2002ff}. 
For heavier Kaluza-Klein modes of gravitino, 
we find for $m_n \ll k$ and $m_n e^{Rk\pi}\gg k$
\begin{eqnarray}
  m_n \approx 6w_0 k, ~
\left( n + \frac{1}{4} \right)\pi ke^{-Rk\pi} 
  \label{ginoh1}
\end{eqnarray}
and for $m_n \gg k$
\begin{eqnarray}
 m_n \approx \left( n - \frac{6w_0}{2\pi} \right)\pi k e^{-Rk\pi}
  \label{ginoh2} .
\end{eqnarray}

\subsection{Hyperscalar Kaluza-Klein mass}

We consider
$n$-th Kaluza-Klein effective field 
$\phi_n^I(x)$ with its mode functions $b_n^I(y)$ as $\phi(x, y)$ 
component and $b_n^{cI}(y)$ as $\phi^c(x, y)$ component 
\begin{eqnarray}
 \left(
  \begin{array}{c}
   \phi(x,y)\\ \phi^c(x,y)
  \end{array}\right)
  =\sum_{n}
\sum_{I=1,2}
 \phi_n^I(x) 
  \left(
   \begin{array}{c}
    b_n^I
(y)\\ \hat \epsilon(y)b_n^c{}^I
(y)
   \end{array}\right), 
\label{modeexpand}
\end{eqnarray}
where $I$ is the indices corresponding to the two independent 
effective fields eigenvalues. 
As for concerning this subsection,
we take the constant superpotential as
\begin{eqnarray}
W_b %\equiv
= 2M_5^3 (w_0 \delta(y) + w_\pi \delta(y-\pi)), 
\label{eq:boundary_pot}
\end{eqnarray}
where $w_{0},w_{\pi}$ are dimensionless constants which 
are assumed to be ${\cal O}(1)$.
After solving the equations of motion for the 
scalar component 
 fields $\phi$ and $\phi^c$,
we find the following Kaluza-Klein mass:
for $\phi^{I=1}$,
\begin{eqnarray} 
 m_n 
\approx k e^{-Rk\pi}
  \left[ \left(n +{2\alpha+1\over 4} \right)\pi 
  \pm \frac{|w_\pi|}{2\sqrt{3}} \right], 
\label{constdx1}
\end{eqnarray}
where the plus (minus) sign should be taken for 
$1/2\le c \le 1 (c \le -1/2~\textrm{or}~c>1)$,
\begin{eqnarray}
\label{phimass}
 m_n &\approx& k e^{-Rk\pi}
  \left[ \left(n +{2\alpha+1\over 4} \right)\pi 
 \right.
\nonumber
\\
&&\left.
 +{w_\pi^2+12\over 24\tan c\pi}
   \left(1-\sqrt{1+{w_\pi^2\over 3}\tan^2 c\pi}\right) 
 \right].  
\end{eqnarray}
for $|c|<1/2$; for $\phi^{I=2}$,
\begin{eqnarray}
 m_n 
\approx k e^{-Rk\pi}
  \left[ \left(n +{2\alpha+1\over 4} \right)\pi 
  \pm \frac{|w_\pi|}{2\sqrt{3}} \right], 
\label{constdx1}
\end{eqnarray}
where the plus (minus) sign should be taken for 
$1/2\le c \le 1 (c \le -1/2~\textrm{or}~c>1)$,
\begin{eqnarray}
\label{phimass}
 m_n &\approx& k e^{-Rk\pi}
  \left[ \left(n +{2\alpha+1\over 4} \right)\pi 
\right.
\nonumber
\\
&&+\left.
{w_\pi^2+12\over 24\tan c\pi}
   \left(1-\sqrt{1+{w_\pi^2\over 3}\tan^2 c\pi}\right) 
 \right].  
\end{eqnarray}
for $|c|<1/2$.
Here $\alpha\equiv |c+1/2|$ and 
$\beta\equiv |c-1/2|$.
The mass splitting depends on the bulk mass parameter for
$|c|<1/2$, which is a new pattern of supersymmetry breaking.

\subsection{Conclusion}

For $w_0\sim 10^7\textrm{GeV}/k$,
$M_5\sim (M_4^2 k)^{1/3}$ and
$c=c_\textrm{\scriptsize cr}-\Delta c$ with
$c_\textrm{\scriptsize cr}\simeq 0.546$ and $\Delta c\sim 10^{-6}$
the orders of various masses are tabulated in Table~\ref{tab:1} (the 
radius is stabilized at $R\sim k^{-1}$).
\begin{table}[htbp]
\caption{A typical mass spectrum}
\label{tab:1}       
\begin{tabular}{ccccc}
\hline\noalign{\smallskip}
Soft & Gravitino & Radion & Moduli & Hyperscalar \\
\noalign{\smallskip}\hline\noalign{\smallskip}
100GeV & $10^7$GeV & 1 TeV & 100TeV & $10^{-2} k$ \\
\noalign{\smallskip}\hline            
\end{tabular}
\end{table}

\noindent{\bf Acknowledgments:}
This work is supported in part by Grant-in-Aid for
Scientific Research from the Ministry of Education,
Culture, Sports, Science and Technology, Japan No.17540237 (N.~S.)
and No.18204024 (N.~M. and N.~S.).
The work of NU is supported by Bilateral exchange program between 
Japan Society for the Promotion of Science and the Academy of Finland.

%
% BibTeX users please use
% \bibliographystyle{}
% \bibliography{}

\end{document}